\documentstyle[11pt]{article} 
\input epsf

\newcommand{\be}{\begin{equation}}
\newcommand{\ee}{\end{equation}}
\newtheorem{theorem}{Theorem}          
\newtheorem{lemma}{Lemma}              

\newtheorem{definition}{Definition}

\newcommand{\hilbert}{{\cal H}}
\newcommand{\fock}{{\cal F}}
\newcommand{\algebra}{{\cal A}}
\newcommand{\cluster}{{\cal C}}
\newcommand{\macros}{{\cal M}}
\newcommand{\tint}{{\cal I}}
\newcommand{\qint}{{\cal T}}
\newcommand{\setg}{{\cal G}}
\newcommand{\Oc}{{\cal O}}
\newcommand{\calh}{{\cal Z}}
\newcommand{\znum}{{\sf Z \!\!\! Z}}
\newcommand{\cnum}{{\sf C \!\!\! C}}

\newcommand{\Sig}{{\sf \Sigma \!\!\! \Sigma}}
\newcommand{\bea}{\begin{eqnarray}} 
\newcommand{\eea}{\end{eqnarray}}
\newtheorem{algo}{Algorithm} 
\begin{document}

\begin{center}
{\LARGE{\bf Suppression of the negative sign problem in quantum Monte Carlo.}}\\[1cm]

\vspace*{1cm}
{\Large A. Galli\footnote{galli@mppmu.mpg.de}}\\[0.3cm]
{\small \em Max-Planck-Institut f\"ur Physik, D-80805 Munich, Germany}\\[2cm]

{\bf Abstract}\\[0.5cm]
\end{center}{\small
We present a new Monte Carlo algorithm for simulating quantum spin systems which is able to suppress the negative sign problem.  This algorithm has only a linear complexity in the lattice size used for the simulation. A general description and a rigorous 
proof of its correctness is given. 
Its efficiency is tested on a simple 2-dimensional fermionic model. 
For this model we show that our algorithm eliminates the sign problem.\\[2cm]
}

\newpage
\section{Introduction}
The study of lattice quantum spin systems is a difficult problem. 
These systems are important in condensed matter physics since they can describe strongly correlated electrons. At present, 
the domain of applicability of the existing analytical methods able to analyze them is still very restricted. To overcome this difficulty numerous groups have tried to study these models using computational techniques.  \\
  
Two major techniques are currently used to study quantum spin systems numerically. The first is the exact diagonalisation method and the second  is quantum Monte Carlo (for a review see \cite{rev}). Usually, both methods have a complexity which grows expo
nentially with the size of the system under study, so that meaningful results are difficult to obtain. In particular, quantum Monte Carlo can suffer from the negative sign problem, which becomes exponentially serious on large lattices and at small tempera
ture. \\                                        
In this work we develop a new algorithm which is able to suppress the sign problem in a quantum Monte Carlo simulation. A general description of this algorithm and a rigorous proof of its correctness are presented. This algorithm was briefly presented in 
\cite{letter}. It is based on a well-controlled approximation which has a linear complexity in the lattice size.\\
 
We have applied it to a simple 2-dimensional models with fermions for  
testing its efficiency.
This model has a serious sign problem if simulated with conventional quantum Monte Carlo algorithms, but if simulated with
our algorithm the sign problem is completely eliminated.
The generality of this algorithm allow us to apply it to any model, opening
new perspectives in the numerical study of quantum spin systems.\\ 
  
This paper is organized as following: first we define the formalism and prove some lemmas (section 2 and 3) that we need for the description of the algorithm. The algorithm is then discussed and a proof of correctness is given (section 3). Finally in sect
ion 4 the results of the example are presented.

\section{Formalism for quantum spin systems on a finite lattice}

We consider a d-dimensional finite lattice $\Lambda\subset  \znum^d$. 
Associated to each site of the lattice we consider a particle with a finite
number $M$ of internal degrees of freedom. A Hilbert space $\hilbert_x$ is
associated with each site $x\in\Lambda$ of the lattice and is isomorphic to
$\cnum^M$. We choose an orthonormal basis $\{e_\sigma^x\}_{\sigma\in I}$ with 
$I=\{1,...,M\}$ in the Hilbert space $\hilbert_x$.
The Hilbert space $\hilbert_\Lambda$ over the lattice $\Lambda$ is given by the tensor product
\be
\hilbert_\Lambda=\bigotimes_{x\in\Lambda}\hilbert_x
\ee
The set $\Omega_\Lambda$ of all configurations ${\omega_\Lambda}$ in $\Lambda$ is defined as the assignment of an element $\sigma_x$ to each lattice site $x\in\Lambda$. The set $\{e_{\omega_\Lambda}\}_{\omega_\lambda\in\Omega_\Lambda}$ 
of all vectors 
\be
e_{\omega_\Lambda}=\bigotimes_{x\in\Lambda}e_{\sigma_x}^x
\ee
forms an orthonormal basis of $\hilbert_\Lambda$.\\

A Fock space is constructed to be able to incorporate the statistics of the particles
\be
\fock_P(\hilbert_\Lambda)=P\bigoplus_{n\geq 0}\hilbert_\Lambda^n\label{fock}
\ee
where $\hilbert_\Lambda^n$ is the n-fold tensor product with itself, 
$\hilbert_\Lambda^0=\cnum$ and $P$ is the projection onto the symmetric or antisymmetric subspace for bosons or fermions, respectively.
\bea
&&P_{Bose}(e_{\omega_\Lambda})=\frac{1}{n!}\sum_\pi 
\bigotimes_{x\in\Lambda}e_{\sigma_\pi(x)}^{\pi(x)}\nonumber\\
&&P_{Fermi}(e_{\omega_\Lambda})=\frac{1}{n!}\sum_\pi 
sgn(\pi)\bigotimes_{x\in\Lambda}e_{\sigma_\pi(x)}^{\pi(x)}
\eea
where $\pi$ is a permutation of the order of the points $(x,\sigma)\in\Lambda\times I$ and 
$sgn(\pi)$ is $+1$ if the permutation $\pi$ is even $-1$ if it is odd. 
We choose an arbitrary order of the points in $\Lambda$ denoted by $x\prec x'$ if $x$ proceeds $x'$ and we define the order of the points in $\Lambda\times I$ by $(x,\sigma)\preceq (x',\sigma')$ if $x\prec x'$ or if $x=x'$, $\sigma\leq \sigma'$.\\

The annihilation and creation operators on $\fock_P(\hilbert_\Lambda)$ are defined as 
\be
c_{x\sigma}=Pa_{x\sigma}P\label{ann}
\ee
and
\be
c^*_{x\sigma}=Pa^*_{x\sigma}P\label{cre}
\ee
with
\bea
&&a_{x\sigma}(e_{\sigma_1}^{x_1}\otimes...\otimes e_{\sigma_n}^{x_n}):=
\sqrt{n}(e_{\sigma}^{x},e_{\sigma_1}^{x_1})e_{\sigma_2}^{x_2}\otimes...\otimes e_{\sigma_n}^{x_n}\nonumber\\
&&a^*_{x\sigma}(e_{\sigma_1}^{x_1}\otimes...\otimes e_{\sigma_n}^{x_n}):=
\sqrt{n+1}\,e_{\sigma}^{x}\otimes e_{\sigma_1}^{x_1}\otimes...\otimes e_{\sigma_n}^{x_n}
\eea
where $(e_{\sigma}^{x},e_{\sigma_1}^{x_1})$ denotes the scalar product of the vectors $e_{\sigma}^{x}$ and $e_{\sigma_1}^{x_1}$. Furthermore, we have 
$a_{x\sigma}|0>:=0$ and $a_{x\sigma}^*|0>=e^x_\sigma$ where $|0>$ denotes the zero particle state (vacuum).\\

For bosons and fermions the operators defined through (\ref{ann}) and (\ref{cre}) satisfy the canonical commutation relations. For bosons we have
\bea
&&[c_{x\sigma},c_{x'\sigma'}]=[c^*_{x\sigma},c^*_{x'\sigma'}]=0\nonumber\\
&&[c_{x\sigma},c^*_{x'\sigma'}]=(e_\sigma^x,e_{\sigma'}^{x'})
\eea
and for fermions
\bea
&&\{c_{x\sigma},c_{x'\sigma'}\}=\{c^*_{x\sigma},c^*_{x'\sigma'}\}=0\nonumber\\
&&\{c_{x\sigma},c^*_{x'\sigma'}\}=(e_\sigma^x,e_{\sigma'}^{x'})
\eea
An orthonormal basis of $\fock_P(\hilbert_\Lambda)$ is given by the vector
\bea
|n_{x_1\sigma_1}...n_{x_k\sigma_k}>=&&
\left(\bigotimes_{q_1=1}^{n_{x_1\sigma_1}}e_{\sigma_1}^{x_1}\otimes...
\bigotimes_{q_k=1}^{n_{x_k\sigma_k}}e_{\sigma_k}^{x_1}\right)_{\prec}\label{bv}
=\nonumber\\
&&=\frac{1}{\sqrt{\prod_{i=1}^k n_{x_i\sigma_i}!}}
\left((c_{x_1\sigma_1}^*)^{n_{x_1\sigma_1}}...(c_{x_k\sigma_k}^*)^
{n_{x_k\sigma_k}}\right)_{\prec}|0>
\eea
where the $(...)_\prec$ denotes that subscripts $(x_i\sigma_i)$ of the braced factors are ordered as discussed above.\\

We define the algebra of observables $\algebra_\Lambda$ 
as the algebra of operators generated by the identity and all monomials of even degree in the creation and annihilation operators associated with the lattice sites $x\in\Lambda$. Two algebras of observables $\algebra_{\Lambda_1}$ and $\algebra_{\Lambda_2}
$ satisfy locality 
\be
\forall A_1\in\algebra_{\Lambda_1},A_2\in\algebra_{\Lambda_2}:
[A_1,A_2]=0 \mbox{ if } \Lambda_1\cap \Lambda_2 = \emptyset
\ee
and inclusion
\be
\algebra_{\Lambda_1}\subseteq  \algebra_{\Lambda_2}\mbox{ if } \Lambda_1 \subseteq \Lambda_2
\ee
The trace of an operator $A\in\algebra_\Lambda$ is defined by
$Tr A =\sum_{v}<v|A|v>$
where $\{|v>\}$ is an orthonormal basis of $\fock_P(\hilbert_\Lambda)$ of the form (\ref{bv}).\\

The dynamics of the system is described by a finite range Hamiltonian 
\be
H=\sum_{B\subset\Lambda} \Phi_B\label{hamiltonian}
\ee 
where $\Phi_B$ is a selfadjoint operator belonging to $\algebra_B$ defined on a bond $B\subset \Lambda$. In our work we consider only periodic boundary 
conditions. The definition of a bond $B$ is then adapted to our boundary conditions.
If the basis vectors (\ref{bv}) are eigenstates of 
$H$ then the Hamiltonian describes a {\em classical} spin system, otherwise it describes a {\em quantum} spin system.
The partition function of the statistical system is given by
\be
Z=Tr e^{-\beta H}\label{Z}
\ee
where $\beta$ is the inverse temperature.
The expectation value of an observable $A\in\algebra_\Lambda$ is obtained by
\be
<A>=\frac{1}{Z}Tr A e^{-\beta H}
\ee
 
\section{Monte Carlo of quantum spin systems}

\subsection{The Trotter formula}

Classical spin systems can be simulated in d-dimensional lattice using Monte Carlo algorithms. 
The weight 
$<{v}|e^{-\beta H}|{v}>$ contributing to the partition function can be directly evaluated
since the basis vector (\ref{bv}) is an eigenvector of the Hamiltonian.
The evaluation of the weights is essential for any Monte Carlo algorithm.\\

For a quantum spin system a Monte Carlo simulation is not straightforward because the basis vector (\ref{bv}) is not an eigenvector of the Hamiltonian any more and the evaluation of the exponential in the partition function becomes a very complex task. In
 fact, if one wants to perform any calculation directly using the d-dimensional lattice state $|v>$, the request in storage and computer time grows like the dimension of the Fock space, which is exponential in the volume $|\Lambda|$ of the lattice. The on
ly practical way to evaluate the weight 
  $<v|e^{-\beta H}|{v}>$ is to apply the Trotter formula 
in a (d+1)-dimensional lattice, where the extra dimension is the discretisation of the inverse temperature (we call it time)
\be
Z=\lim_{T\rightarrow\infty}Tr \left(e^{-\beta \epsilon H}\right)^T
\label{trotter}
\ee
where $\epsilon =1/T$. For a simulation we have to restrict the interval 
$\tint=[0,T]\cap\znum$ to a finite set, approximating in this way the partition function. The systematic errors introduced by the approximation are of order $\epsilon^2$. 

We consider a decomposition of the Hamiltonian (\ref{hamiltonian}) 
into a sum of $q$ operators
\be
H=\sum_{i=1}^q H_i\label{hq}
\ee
where each of them is a sum of commuting selfadjoint operators
\be
H_i=\sum_{B_i}\Phi^i_{B_i}\label{dec}
\ee
From the locality of the algebra of operators we can construct this sum so that the operators $\Phi^i_B$ satisfy 
$$[\Phi^i_{B_i},\Phi^{i'}_{B'_{i'}}]=0\mbox{ if } B_i\cap B'_{i'} =\emptyset 
$$
This means that for each label $i=1,...q$ we consider the sum $ \sum_{B_i}$ over disjoint bonds.\\

We can define a classical (d+1)-dimensional spin system starting from the Trotter formula and inserting a projector 
\be
{\bf 1}_{(i,t)}=\sum_{v_i(t)}|v_i(t)><v_i(t)|
\ee
for each time point in $t\in\tint$ and each exponential 
$e^{-\beta\epsilon H_i}$.
Here the sum $\sum_{v_i(t)}$ goes over an orthonormal basis of
 $\fock_P(\hilbert_\Lambda)$. 
The partition function is then approximated by
\bea
Z\simeq &&\sum_{v(0)}<v(0)|\left[\prod_{t=1}^T
\left(\prod_{i=1}^q e^{-\beta \epsilon H_i} \right)
\right]|v(0)>=\nonumber\\
=&&
\sum_{v}<v(0)|
\left[\prod_{t=1}^T
\left(\prod_{i=1}^q e^{-\beta \epsilon H_i}|v_i(t)>
<v_i(t)| \right)
\right]|v(0)>\label{trotta}
\eea
Here the sum $\sum_{v(0)}$ goes over an orthonormal basis of  $\fock_P(\hilbert_{\Lambda})$.
The sum $$\sum_{v}:=\sum_{v_1(0),...,v_q(0),...,v_1(T),...,v_q(T)}$$ 
goes over an orthonormal basis of
 $\fock_P^\qint(\hilbert_{\Lambda})$ where 
$\qint=[0,q\times T]\cap\znum$ and
\be
\fock_P^\qint(\hilbert_{\Lambda}):=\bigotimes_{\tau\in\qint} \fock_P(\hilbert_{\Lambda})
\ee
replicates the Fock space $\fock_P(\hilbert_{\Lambda})$ at each slice $\tau\in \qint$.
Using the locality of the algebra of observables and the decomposition of the Hamiltonian (\ref{dec}) in commuting operators we can factorize the contributions in (\ref{trotta}) in a product of terms localized over the bonds $B_i$
\be
Z\simeq \sum_v w(v)\label{weight}
\ee
where the weight of the vector 
$|v>\in \fock_P^\qint(\hilbert_{\Lambda})$ is given by the product of terms
\be
w(v)=\prod_{t=0}^T\label{weightdef}
\prod_{i=1}^q \prod_{B_i}w_{B_i}(v,(i,t))
\ee
with, for $t>0$,
\be
w_{B_i}(v,(i,t))=<v_i(t)|e^{-\beta \epsilon \Phi^{i'}_{B_{i'}}}|v_{i'}(t')>
\label{weightlocal}
\ee
and
\be
\left\{
\begin{array}{lll}
t'=t,&i'=i+1 &\mbox{ if } 1\leq i<q\\
t'=t+1,&i'=1 &\mbox{ otherwise}
\end{array}
\right.
\ee
and, for $t=0$,
\be
w_{B_i}(v,(i,0))=
\left\{
\begin{array}{lr}
<v(0)|e^{-\beta \epsilon \Phi^{1}_{B_{1}}}|v_{1}(1)> & \mbox{if } i=1\\
1&\mbox{otherwise}
\end{array}
\right.
\ee
The weight $w_{B_i}(v,(i,t))$ is a real number. 
We can decompose the weight $w_{B_i}(v,(i,t))$ in its modulus and its sign.
\be
w_{B_i}(v,(i,t))=sgn(w_{B_i}(v,(i,t)))\times |w_{B_i}(v,(i,t))|
\label{weightsign}
\ee
The evaluation of the weight $|w_{B_i}(v,(i,t))|$ requires to perform the exponential $e^{-\beta\epsilon \Phi^{i'}_{B_{i'}}}$ of the operator $\Phi^{i'}_{B_{i'}}$. Because this operator is local, it is a finite matrix 
of dimension equal to the dimension of the Fock space localized on the bond $B_i$. The computational complexity of this operation is then very small.
The sign of the weight $sgn(w_{B_i}(v,(i,t)))$ is not generally a local quantity. For fermionic systems its evaluation requires the calculation of the sign of the permutation $\pi$ associated to the order in (\ref{bv}) of the $<v_i(t)|$ and $|v_{i'}(t')>$
 vectors. Also this operation is relatively easy. For bosonic systems the sign can also be negative but its evaluation remains local.

\subsection{Example}

As an example we consider a simple model of free quantum spin systems\footnote{This example was analyzed by quantum Monte Carlo in \cite{wiese}.}. 
We consider fermions living on the sites of a spatially 2-dimensional
square lattice with periodic boundary conditions. 
We consider a particle with only one internal degree of freedom $I=\{1\}$.
We chose $\{e_\sigma^x\}_{\sigma\in I}$ to be an orthonormal basis of the Hilbert space $\hilbert_x$.
The Fock space is constructed as in (\ref{fock}). Creation and annihilation 
operators
$c_x^*$ and $c_x$ anticommute
\begin{equation}
\{c_x^*,c_y^*\} = 0, \,\,\, \{c_x,c_y\} = 0, \,\,\,
\{c_x^*,c_y\} = \delta_{xy}.
\label{anticommutators}
\end{equation}
for $x,y\in\Lambda$.
We
consider the Hamilton operator
\begin{equation}
H = \sum_{x,i} (c_x^* c_x + c_{x+\hat{i}}^* c_{x+\hat{i}}
- c_x^* c_{x+\hat{i}} - c_{x+\hat{i}}^* c_x),
\end{equation}
where $\hat{i}$ is the unit vector in the $i$-direction. The model is trivial and
can be solved in momentum space. However, when one tries to simulate it with a Monte Carlo algorithm it shows from the algorithmic point of view all the 
characteristic features of more complicated quantum spin systems.
We can approximate the partition function $Z$ with the Trotter formula (\ref{trotta}) by decomposing the Hamiltonian into four pieces $H = H_1 + H_2 + H_3 + H_4$
(see eq. (\ref{hq})) and 
\begin{equation}
H_1 = \sum_{x=(2m,n)} h_{x,1}, \,\,\, H_2 = \sum_{x=(m,2n)} h_{x,2}, \,\,\,
H_3 = \sum_{x=(2m+1,n)} h_{x,1}, \,\,\, H_4 = \sum_{x=(m,2n+1)} h_{x,2},
\end{equation}
where $h_{x,i} = c_x^* c_x + c_{x+\hat{i}}^* c_{x+\hat{i}}
- c_x^* c_{x+\hat{i}} - c_{x+\hat{i}}^* c_x$ (see eq. (\ref{dec})).
The individual contributions to a given $H_j$ commute with each other,
but two different $H_j$ do not commute. 

Using (\ref{trotta}) we can write the grand canonical partition function
\be
Z = Tr e^{- \beta (H-\mu N)}
= lim_{T \rightarrow \infty} Tr
[
e^{- \beta \epsilon (H_1-\frac{\mu}{4}N) } 
e^{- \beta \epsilon (H_2-\frac{\mu}{4}N) } 
e^{- \beta \epsilon (H_3-\frac{\mu}{4}N) } 
e^{- \beta \epsilon (H_4-\frac{\mu}{4}N) } 
]^T
\ee
where $\mu$ is the chemical potential and $N=\sum_{x\in\Lambda}n_x$ with $n_x=c^*_xc_x$ is the particle number operator.
Inserting a complete set of Fock states between the factors
and using the locality of $h_{x,i}$ for bonds $<x,x+\hat i>$ 
we obtain matrix elements of the exponential (\ref{weightlocal}).
\bea
&&e^{- \beta \epsilon (h_{x,i}-\frac{\mu}{4}(n_x+n_{x+\hat i}))} =
e^{- \beta \epsilon(1-\frac{\mu}{4})} \times \nonumber\\&&
\left(\begin{array}{cccc}
\exp(\beta\epsilon(1-\frac{\mu}{4})) & 0 & 0 & 0 \\
0 & \cosh(\beta\epsilon)  & \sinh(\beta\epsilon) \Sig & 0 \\
0 & \sinh(\beta\epsilon) \Sig & \cosh(\beta\epsilon)  & 0 \\
0 & 0 & 0 & \exp(-\beta\epsilon(1-\frac{\mu}{4})) \end{array} \right)
\label{matrix}
\eea
where the $4 \times 4$ matrix is in the Fock space basis $|00>$,
$|01>$, $|10>$, $|11>$ defined on the bond $<x,x+\hat{i}>$.
$\Sig $ is the sign of the weight which has to be evaluated looking at the permutations of the fermions with respect to the given order of the basis vectors (\ref{bv}). The partition function is approximated by a classical spin system over occupation numb
ers
$n(x,\tau) = 0,1$  (here $\tau$ labels the $T\cdot q$ time slices in $\qint$ with $q=4$)
with periodic boundary conditions. The systematic errors due to the finite T
are of order $\epsilon^2$.
\begin{equation}
Z \simeq \prod_{x,\tau} \sum_{n(x,\tau) = 0,1} |w(n)| \, sgn(w(n)) .
\end{equation}
The modulus factor takes the form
\begin{eqnarray}
&&|w(n)| =\!\!\! \prod_{x=(2m,n),\tau=4p}
w[n(x,\tau),n(x+\hat{1},\tau),n(x,\tau+1),n(x+\hat{1},\tau+1)]
\nonumber \\ && \times
\prod_{x=(m,2n),\tau=4p+1}
w[n(x,\tau),n(x+\hat{2},\tau),n(x,\tau+1),n(x+\hat{2},\tau+1)]
\nonumber \\ && \times
\prod_{x=(2m+1,n),\tau=4p+2}
w[n(x,\tau),n(x+\hat{1},\tau),n(x,\tau+1),n(x+\hat{1},\tau+1)]
\nonumber \\ && \times
\prod_{x=(m,2n+1),\tau=4p+3}
w[n(x,\tau),n(x+\hat{2},\tau),n(x,\tau+1),n(x+\hat{2},\tau+1)], \nonumber \\
\end{eqnarray}
with $w[0,0,0,0] = 1$, $w[1,1,1,1] = e^{-2 \beta\epsilon(1-\frac{\mu}{4})}$,
$w[0,1,0,1] = w[1,0,1,0] = e^{-\beta \epsilon(1-\frac{\mu}{4})}\cosh(\beta\epsilon)$,
$w[0,1,1,0] = w[1,0,0,1] = e^{-\beta \epsilon(1-\frac{\mu}{4})} \sinh(\beta\epsilon)$.
All other values are zero. 
The occupation numbers
$n(x,\tau) = 0,1$ are variables interacting with each other via the
time-like plaquette couplings
$w[n(x,\tau),n(x+\hat{i},\tau),n(x,\tau+1),n(x+\hat{i},\tau+1)]$.

Each state is weighted by a sign factor which arises 
from the fermionic statistics.
The sign factor $sgn[w(n)]$ is a product of terms
$sgn[n(x,\tau),n(x+\hat{i},\tau),n(x,\tau+1),n(x+\hat{i},\tau+1)]$ associated with
each plaquette interaction. One has
$sgn[0,0,0,0]$ $= sgn[1,1,1,1]$ $=
sgn[0,1,0,1]$ $= sgn[1,0,1,0] = 1$. A nontrivial
sign $\pm 1$ may arise only for plaquette interactions of type $[0,1,1,0]$ and
$[1,0,0,1]$.

\subsection{The sign problem}
The decomposition of the Hamiltonian in local terms (\ref{hq},\ref{dec}) allows us to treat the new (d+1)-dimensional spin system as a classical spin system with state space $\fock_P^\qint(\hilbert_\Lambda)$,
 partition function (\ref{weight}) and weight (\ref{weightdef}). 
The evaluation of the weight is of small complexity, contrary to the original d-dimensional quantum spin system (\ref{Z}). 
The price to pay is that the new classical system has a partition function (\ref{weight}) which has not generally positive semidefinite weights $w(v)$. This fundamental difficulty is usually referred to as the "negative sign" problem. It is not related to
 any approximations in the Monte Carlo scheme but it describes the fact that the statistical error of the observables can become very large, increasing exponentially in the inverse temperature $\beta$ and lattice volume $|\Lambda\times \qint|$.\\

\begin{definition}{\rm 
A {\em classical observable} is an operator $A\in\algebra_\Lambda$ acting on the Fock space $\fock_P(\hilbert_\Lambda)$ defined at $t=0$ and diagonal with respect to the basis of the form (\ref{bv}). We denote by $A(v)$ the matrix element $A(v)=<v(0)|A|v(
0)>$.  }\label{cobs}
\end{definition}

For simplicity we restrict our discussion to classical observables. With some minor modifications our algorithm is applicable also for observables which are not diagonal with respect to the basis (\ref{bv}). We emphasize that, usually, the interesting obs
ervables are calssical.\\

The expectation value of a classical observable can be written
\be
<A>=\frac{\sum_v A(v)\,w(v)}{\sum_v w(v)}= \frac{\sum_v A(v)\,|w(v)|sgn(w(v))}{\sum_v |w(v)|sgn(w(v))} \label{ev}
\ee
A Monte Carlo algorithm needs a positive semidefinite weight to be able to construct a Markov process. Redefining the observable incorporating the sign in it 
$
\tilde A(v)=sgn(w(v))A(v)
$ we can simulate a positive semidefinite classical spin system with statistical weight $|w(v)|$ and correct the measurement of the observable
\be
<A>_w= \frac{\sum_v \tilde A(v)\,|w(v)|}{\sum_v |w(v)|}\times \frac{1}{<sgn(v)>_{|w|}}
=:\frac{<A\,sgn>_{|w|}}{<sgn>_{|w|}}\label{evplus}
\ee
where $<...>_w$ and $<...>_{|w|}$ denote the averages taken with the weights $w$ and $|w(v)|$, respectively. If the average sign is small there will be large cancelations making an accurate evaluation of $<A>$ impossible.\\

\subsection{Equivalent Monte Carlo algorithms}

We consider a Monte Carlo algorithm $\phi_{|w|}$ which produces a Markov chain with state space $\{|v>\in\fock_P^\qint(\hilbert_{\Lambda})\}$ and equilibrium distribution $|w(v)|$. 
We assume that this algorithm generates a state $|v'>$ from a state $|v>$
with a transition probability $T(v'\leftarrow v)$
\be
\phi_{|w|}:|v>\rightarrow |v'>
\ee 
We assume stationarity 
\be
\sum_v T(v'\leftarrow v)|w(v)|=|w(v')|\label{stationarity}
\ee
and irreducibility (ergodicity\footnote{Notice that only on Markov chains with {\em finite} state space, "ergodic" is used in the physics literature as a synonym for "irreducible" (see \cite{ergodic1}, Section 2.4). On Markov chains with general state spa
ce they have a different meaning (\cite{ergodic3}, p.169).})
\be
T(v'\leftarrow v)>0,\,\,\,\,\forall |v>,|v'>\in \fock_P^\qint(\hilbert_{\Lambda})\mbox{ with $|w(v)|>0$ and $|w(v')|>0$}\label{irreducibility}
\ee
to ensure the convergence to equilibrium.
This algorithm can be realized in different ways. The explicit realization of it is for the moment irrelevant to the discussion. We denote the expectation value measured with the algorithm $\phi_{|w|}$ by eq. (\ref{evplus}).\\

If the average of the sign is small this algorithm will suffer from the sign problem as discussed above.
It is important to notice that the value $A(v)$ of a classical observable $A$ does not depend on the complete state $|v>$ but only on the part living on the time slice at $t=0$. This is a crucial property that we will use for constructing an algorithm whi
ch is able to reduce the sign problem by substantially increasing the average sign. We first introduce some simple definitions and lemmas which allow us to construct this algorithm.\\

\begin{definition}{\rm 
A mapping 
\bea
g:&&\fock_P^\qint(\hilbert_{\Lambda})\rightarrow \fock_P^\qint(\hilbert_{\Lambda})\nonumber\\
&&|v>\mapsto g|v>
\eea
is called {\em observable preserving} if for any 
classical observable $A$ its value $A(v)$ is $g$ invariant $A(v)=A(gv)$. 
Such a mapping is easily constructed: we can require that it does not change the state $|v>$ at $t=0$. } 
\label{preserving}
\end{definition} 

\begin{lemma}{\rm
We consider a set $\setg$ of observable preserving and bijective mappings $g$.  We define the new weight 
\be
\tilde w(v,g)=\frac{1}{2}(w(v)+w(gv))\label{tildew}
\ee
then we obtain for the expectation value (\ref{ev}) of a classical observable 
\be
\frac{\sum_v A(v)w(v)}{\sum_v w(v)}=
\frac{\sum_{g\in\setg}\sum_v A(v)\tilde w(v,g)}{\sum_{g\in\setg}\sum_v \tilde w(v,g)}
\label{newev}
\ee
}
\label{lemmanewev}
\end{lemma}
{\small \underline{Proof}: Using the bijective property of $g$, it is easy to see that after summing over all states $|v>\in
\fock_P^\qint(\hilbert_{\Lambda})$ 
we obtain
$\sum_v w(v)=\sum_v w(gv)$. We denote by $|\setg|$ the number of elements in $\setg$ and after summation over all mappings $g\in \setg$ we obtain that
$\sum_v w(v)=\frac{1}{|\setg|}\sum_{g\in\setg}\sum_v \tilde w(v,g)$.
Using the fact that $g$ is observable preserving (that means $A(v)=A(gv)$), 
in analogy as before we can see that 
$
\sum_v A(v) w(v)=\frac{1}{|\setg|}\sum_{g\in\setg}\sum_v A(v)\tilde w(v,g) 
$ so that (\ref{newev}) holds.$\Box$} 
\begin{definition}{\rm 
We call two Monte Carlo algorithms {\em equivalent} if for all classical observables $A$ their expectation values measured with both algorithms are equal.
We call two Monte Carlo algorithms {\em $\delta$-quasi equivalent} if for all classical observables $A$ their expectation values measured with both algorithms are equal up to systematic errors of order $\delta$.}
\end{definition}
Using this last lemma \ref{lemmanewev} it is clear that the expectation value of $A$ measured with a Monte Carlo algorithm $\phi_{|w|}$ is the same as the one measured with a Monte Carlo algorithm $\phi_{|\tilde w|}$ so that 
$\phi_{|w|}$ and $\phi_{|\tilde w|}$ are equivalent. 
The set of mappings $\setg$ can contain an arbitrary number of elements. 
We suppose that we can construct a mapping $g$ so that the average of the sign if measured with $\phi_{|\tilde w|}$ and $\tilde w$ defined as in (\ref{tildew}) satisfies 
\be
<sgn(v)>_{|\tilde w|}\,\,\,\,\geq\,\,\,\,<sgn(v)>_{|w|}
\label{suppression}
\ee
where $<...>_{|\tilde w|}$ and $<...>_{|w|}$ mean the expectation taken from the $|\tilde w|$ and $|w|$ distributions, respectively. 
In the case that the average sign {\em substantially} increases then it is evident that the sign problem is suppressed. 
Notice that the sign is not a classical observable because it is a global quantity depending on the state over the complete lattice $\Lambda\times \qint$. Two equivalent Monte Carlo algorithms can thus have different sign expectation values. The sign prob
lem is now shifted to the search of a mapping with the desired property. Such a mapping can be constructed looking for clusters of spins to be flipped so that (\ref{suppression}) is satisfied with high probability during the Monte Carlo process. 
This is a non trivial problem, since to ensure the bijectivity of the mapping the required work grows exponentially in the volume of the lattice $\Lambda\times \qint$ because a deterministic cluster search is needed. Apart from some trivial example, where
 the amount of work is not too excessive, this approach is not convenient.\\

However, the problem can be solved by constructing a mapping which is bijective with probability $1-\delta$ at each Monte Carlo iteration where $\delta$ is very small. We will see that the construction of such a mapping has a complexity which is linear in
 the volume of the lattice because a randomized cluster search can be used in connection with a hashing technique for testing the bijectivity. Since the mapping is not always bijective, during the Monte Carlo process a systematic error is produced, so tha
t $\phi_{|\tilde w|}$ is only $\delta$-quasi equivalent to $\phi_{|w|}$. If we choose $\delta$ to be smaller than the statistical error and the systematic error $\epsilon^2$ introduced by the Trotter formula, the precision of our algorithm is sufficient.
In the next subsection we will show how this mapping can be constructed explicitly.

\subsection{Reduction of the sign problem}    
We want to construct a mapping $g$ which has the property (\ref{suppression}) and can be used to construct a Monte Carlo algorithm $\delta$-quasi equivalent to $\phi_{|w|}$. First we introduce some concepts which allow us to construct this mapping.

\begin{definition}{\rm 
A mapping 
\bea
g:&&\fock_P^\qint(\hilbert_{\Lambda})\rightarrow \fock_P^\qint(\hilbert_{\Lambda})\nonumber\\
&&|v>\mapsto g|v>
\eea
is called {\em compatible} in $|v>$ if $w(v)\neq 0$ then
$w(gv)\neq 0$.\label{compatible}}
\end{definition}

\begin{definition}{\rm
A state $|v>$ of the Fock space $ \fock_P^\qint(\hilbert_{\Lambda})$ can be rewritten as
\be
|v>=:\bigotimes_{(x,\tau)\in\Lambda\times \qint} |n_\sigma(x,\tau)>
\ee
where $\sigma\in I$ is defined in section 2.
A {\em flip} $\Xi_{(x,\tau)}$ with $(x,\tau)\in\Lambda\times \qint$ is a mapping from $\fock_P^\qint(\hilbert_{\Lambda})$ to $\fock_P^\qint(\hilbert_{\Lambda})$
which locally transforms $|n_\sigma(x,\tau)>$ in some $|n'_{\sigma'}(x,\tau)>$ so that $\Xi_{(x,\tau)}\circ\Xi_{(x,\tau)}=1$.
The composition of flips over a subset $\Omega\subset \Lambda\times \qint$ is denoted by $\Xi_\Omega$.
A subset $\cluster\subset\Lambda\times\qint$ is called a {\em cluster} of $|v>$ if $\Xi_\cluster$ is compatible in $|v>$. A subset $\cluster\subset\Lambda\times\qint$ is called a {\em preserving cluster} of $|v>$ if $\Xi_\cluster$ is compatible in $|v>$ a
nd observable preserving. We call {\em macros} of $|v>$ the set of preserving clusters of a state $|v>$ with respect to a flip $\Xi$ and denote it by $\macros(v,\Xi)$. For a given flip we just denote it by $\macros(v)$ if the context allows that. We defin
e an arbitrary {\em order of clusters} in $\macros(v)$ and denote it by 
$\cluster\prec\cluster'$ if $\cluster$ precedes $\cluster'$. We consider the empty cluster $\emptyset\in\macros(v)$ to be $\emptyset\succ \cluster,\,\,\,\forall \cluster\in\macros(v)$. } \label{clusterdef}
\end{definition} 
We now proceed to the construction of the mapping $g$ with the desired properties.
\begin{algo}{\rm
We consider a hashing function $h$ which assigns in an arbitrary way a non vanishing integer label to any state
\bea
h:&& \fock_P^\qint(\hilbert_{\Lambda})\rightarrow \calh\subset \znum\nonumber\\
&&|v>\mapsto h(v)
\eea
We suppose that we can store a hashing table $H_{table}$ 
with $|\calh|$ integer entries. Here $|\calh|$ denotes the number of different labels and $|\calh|\leq dim (\fock_P^\qint(\hilbert_{\Lambda}))$.
Given a state $|v>$ we define $g|v>$ with the following procedure: Let $\macros'\subset\macros(v)$ be an arbitrary subset of $\macros(v)$ containing the empty cluster $\emptyset$, then{\tt
\bea
&&\mbox{select the first $\cluster\in\macros'$}\nonumber\\
\mbox{repeat} &&\nonumber\\
&&\mbox{if ({\sf condition}={\bf true}) then} \nonumber\\
&&\,\,\,\,\,\mbox{$g|v>=\Xi_\cluster |v>$}\nonumber\\
&&\mbox{otherwise}\nonumber\\ 
&&\,\,\,\,\,\mbox{$g|v>=|v>$}\nonumber\\
&&\,\,\,\,\,\mbox{select next $\cluster\in\macros'$}\nonumber\\
&&\mbox{end if}\nonumber\\
\mbox{until}&&\mbox{({\sf condition}={\bf true}) or ($\cluster=\emptyset$)}\nonumber
\eea}
where we define the  
\be
\mbox{{\sf condition}}=\{(w(v)+w(\Xi_\cluster v)>0)\mbox{ {\bf and} } \Oc(v,\Xi_\cluster v)\}\label{ccc}
\ee
The boolean function $\Oc(v,v')$ is defined by the following procedure
{\tt
\bea
\Oc(v,v'):&&
\mbox{if $H_{table}(h(v'))=0$ then $H_{table}(h(v')):=h(v)$ endif}\nonumber\\
&&\mbox{if $H_{table}(h(v'))\neq h(v)$ then output $\Oc$={\bf false}}\nonumber\\
&&\mbox{else output $\Oc$={\bf true} endif}\nonumber
\eea}
}\label{alg}
\end{algo}
The selection of the clusters in the macros follows the chosen order of the clusters. Practically there is no need to find all the clusters of a macros. One can select a point $(x,\tau)\in\Lambda\times \qint$ and construct a cluster starting from it. Duri
ng the construction a fixed list of random numbers can be used. It is important, however, that this list remains always the same every time one applies this procedure to a state $|v>$. Changing the list of points or the list of random numbers is equivalen
t to select a new mapping $g$ in $\setg$. If the constructed cluster does not satisfy the {\sf condition} (\ref{ccc}) the next point in $\Lambda\times \qint$ can be selected and a new cluster constructed and this search is repeated until the {\sf conditio
n} (\ref{ccc}) is satisfied. If the {\sf condition} (\ref{ccc}) is never satisfied the procedure can be stopped, for example, when all the points in $\Lambda\times \qint$ are tested once, and the original state is returned as the result. In this way the s
earch of the cluster has a linear average complexity in the size of the lattice.\\

\begin{theorem}{\rm
Let $\setg$ a set of mappings defined by the algorithm \ref{alg}.  
A weight $\tilde w$ is defined by
\be
\tilde w(v,g)=\frac{1}{2}(w(v)+w(gv))\label{w2}
\ee
We consider a Monte Carlo algorithm $\phi_{|\tilde w|}$ 
 which produces a Markov chain with state space $\{\psi=(|v>,g)\in\fock_P^\qint(\hilbert_{\Lambda})\times \setg\}$ and equilibrium distribution $|\tilde w(v,g)|$. 
We assume that this algorithm generates a state $\psi'=(|v'>,g')$ from a state $\psi=(|v>,g)$ with a transition probability $T(\psi'\leftarrow \psi)$.
We assume stationarity 
\be
\sum_\psi T(\psi'\leftarrow \psi)|\tilde w(v,g)|=|\tilde w(v',g')|\label{stationarity2}
\ee
and irreducibility 
\be
T(\psi'\leftarrow \psi)>0,\,\,\,\,\forall \psi,\psi'\in \fock_P^\qint(\hilbert_{\Lambda})\times \setg \mbox{ with $|\tilde w(v,g)|>0$ and $|\tilde w(v',g')|>0$}\label{irreducibility2}
\ee
to ensure the convergence to equilibrium.\\
Then this Monte Carlo is $\delta$-quasi equivalent to $\phi_{|w|}$ with $\delta=\frac{1}{|\calh|}$ and 
\be
<sgn(v)>_{|\tilde w|}\,\,\,\,\geq\,\,\,\,<sgn(v)>_{|w|}\label{46}
\ee
}\label{teorema}
\end{theorem}
{\small \underline{Proof}: The flip $\Xi_\cluster$ is defined observable preserving (see definition \ref{clusterdef}) so that all $g$ are also. The boolean function $\Oc$ in the {\sf condition} (\ref{ccc}) guarantees us that the mapping $g$ maps a state $
|v>$ bijectively to a state $|v'>$ with probability $1-O\left(\frac{1}{|\calh|}\right)$. Because of lemma \ref{lemmanewev} it is then clear that $\phi_{|\tilde w|}$ is equivalent to $\phi_{|w|}$ up to errors of order $\frac{1}{|\calh|}$ in the average of 
an observable, so that the two algorithms are $\delta$-quasi equivalent with $\delta=\frac{1}{|\calh|}$. The {\sf condition} (\ref{ccc}) in algorithm \ref{alg} guarantees us that 
$$
sgn(\tilde w(v))=
\left\{\begin{array}{ll}
sgn(w(v)+w(\Xi_\cluster(v)))=1&\mbox{if a cluster $\cluster$ satisfying} \\
                              &\mbox{{\sf condition} is found in $\macros'$,}\\
sgn(w(v))&\mbox{otherwise}
\end{array}\right\}\geq sgn(w(v))
$$
so that after average (\ref{46}) is satisfied. $\Box$}\\[0.5cm]
It is important to notice that the systematic error of order $\delta$ introduced by this algorithm can be tuned by increasing the dimension $|\calh|$ of the hashing table.

\subsection{Monte Carlo algorithm for the weight $\tilde w$.}
A Monte Carlo algorithm for the spin system described by the weight $\tilde w(v,g)$ needs a method for updating the state $\psi$ to a new state $\psi'$ with a transition probability $\tilde T(\psi'\leftarrow \psi)$ which satisfies stationarity (\ref{stati
onarity2}) and irreducibility (\ref{irreducibility2}). The modulus of the weight 
$|\tilde w|$ is not local, contrary to $|w(v)|$ (see section 3.1), because after adding the term $w(\Xi_\cluster v)$ in $\tilde w$ we can not factorize $\tilde w$ anymore in a product like (\ref{weightdef}). To find an efficient updating method applicable
 directly to $\tilde w$ is a difficult task. One can, however, update the states $|v>$ using the Monte Carlo algorithm $\phi_{|w|}$ which is supposed to be known and then correct the weight distribution using an accept/reject global Metropolis test. 

\begin{algo}{\rm \label{alg2}
Suppose we know an Monte Carlo algorithm $\phi_{|w|}$ for the distribution 
$|w(v)|$. For simplicity we suppose that the Monte Carlo algorithm $\phi_{|w|}$ satisfies irreducibility and the detailed balance condition 
\be
T(v'\leftarrow v)|w(v)|=T(v\leftarrow v')|w(v')|
\label{dbal}
\ee
where $T(v'\leftarrow v)$ is the transition probability of $\phi_{|w|}$.
It is clear that stationarity follows from this condition.
We consider a set of mappings $\setg$ constructed as in algorithm \ref{alg}.
We define the Monte Carlo algorithm $\phi_{|\tilde w|}$ for the equilibrium distribution $|\tilde w(v,g)|$ and the state space 
$\{\psi=(|v>,g)\in\fock_P^\qint(\hilbert_{\Lambda})\times \setg\}$
by the procedure
\bea
&&\nonumber\\
&&\mbox{{\tt input the state $\psi=(|v>,g)$}}\nonumber\\
&&\mbox{{\tt generate $|v'>=\phi_{|w|}|v>$ using $\phi_{|w|}$}}\nonumber\\
&&\mbox{{\tt select randomly a mapping $g'\in\setg$}}\nonumber\\
&&P_A(\psi',\psi)=min\left(1,\left[\frac{|\tilde w(v',g')|\cdot|w(v)|}{|\tilde w(v,g)|\cdot|w(v')|}\right]\right)\nonumber\\
&&\mbox{{\tt accept $\psi'=(|v'>,g')$ as the new state with probability $P_A(\psi',\psi)$}}\nonumber
\eea }
\end{algo}
\vspace{0.2cm}
\begin{theorem}{\rm 
Let $\phi_{|\tilde w|}$ be defined as in algorithm \ref{alg2}. Then $\phi_{|\tilde w|}$ satisfies stationarity (\ref{stationarity2}) and irreducibility (\ref{irreducibility2}).}\label{teoremuccio}
\end{theorem}
{\small\underline{Proof}: 
Stationarity can be proven by showing that the detailed balance condition 
$$
\tilde T(\psi'\leftarrow \psi)|\tilde w(\psi)|=\tilde T(\psi\leftarrow \psi')|\tilde w(\psi')|
$$
is satisfied by $\phi_{|\tilde w|}$ where 
$$
\tilde T(\psi'\leftarrow \psi)=T(v'\leftarrow v)\cdot P_A(\psi',\psi)
$$ 
is the transition probability for $\phi_{|\tilde w|}$. We suppose that 
$P_A(\psi',\psi)<1$ for the states $|\psi>$ and $|\psi'>$. Using the property that
$P_A(\psi,\psi')=1$ if  $P_A(\psi',\psi)<1$ and eq. (\ref{dbal}) we have
\bea
\tilde T(\psi'\leftarrow \psi)|\tilde w(\psi)|&&=T(v'\leftarrow v)\cdot P_A(\psi',\psi)
|\tilde w(\psi)|=T(v'\leftarrow v)\frac{|\tilde w(\psi')|\cdot|w(v)|}{|\tilde w(\psi)|\cdot|w(v')|}|\tilde w(\psi)|=\nonumber\\
&&=T(v'\leftarrow v)|w(v)|\frac{|\tilde w(\psi')|}{|w(v')|}=
T(v\leftarrow v')|w(v')|\frac{|\tilde w(\psi')|}{|w(v')|}=
\nonumber\\
&&=T(v\leftarrow v')|\tilde w(\psi')|=T(v\leftarrow v')\cdot P_A(\psi,\psi')|\tilde w(\psi')|=\nonumber\\
&&=\tilde T(\psi\leftarrow \psi')|\tilde w(\psi')|
\nonumber
\eea
The case $P_A(\psi',\psi)=1$ and $P_A(\psi,\psi')<1$ is analogous.
Irreducibility is clear because $\tilde w(\psi)\neq 0$ when $w(v)\neq 0$ so that $P_A(\psi',\psi)>0$ and because $\phi_{|w|}$ satisfies irreducibility. $\Box$}\\[0.8cm]
If the set of mappings $\setg$ contains only one element then the dimension of the hashing table has to be larger than the number of Monte Carlo iterations one desires to perform. In this way one avoids too many collisions in the hashing table. This, of c
ourse, is doable, but requires a lot of memory. If, however, the set of mappings $\setg$ contains a huge amount of elements then the hashing table can be small because the algorithm uses a selected mapping only for a short time and then selects a new one 
according to the acceptance probability $P_A$. If the set $\setg$ is big enough, the probability that the algorithm 
selects the same mapping twice is 
infinitesimal so that there is no need to store the history of the hashing tables of old mappings.\\ 

Theorems \ref{teorema} and \ref{teoremuccio} show that 
a Monte Carlo simulation can be performed using $\phi_{|\tilde w|}$ and a classical observable $A$ can be measured using
\be
<A>_w=\frac{<A\,sgn>_{|\tilde w|}}{<sgn>_{|\tilde w|}}+O\left(\frac{1}{|\calh|}\right)
\ee
If the {\sf condition} (\ref{ccc}) is satisfied with high probability during the Monte Carlo process then the negative sign problem is eliminated. The systematic error $O\left(\frac{1}{|\calh|}\right)$ produced by the hashing technique used for checking t
he bijectivity of the mappings $g$ can be explicitly measured durung the simulation.

\section{Application to the example}
We apply our algorithm \ref{alg2} to the example 
presented in section 3.2. We use for the updating of the states $|v>$ with respect to the distribution $|w(v)|$ a standard loop algorithm \cite{evertz,wiese} which we denote as $\phi_{|w|}$. A description of this algorithm applied to the example of sectio
n 3.2 is given in \cite{wiese}. The reduction of the sign problem is then realized using our algorithm \ref{alg2}.\\ 

The loop algorithm is essentialy a cluster algorithm \cite{cluster} where a cluster $\cluster$ in the macros $\macros(v)$ is selected with a certain probability so that an update of $|v>$ realized by a flip $\Xi_\cluster$ satisfies detailed balance. For c
ompleteness, we briefly describe the loop algorithm we have used in the example.
We define the flip $\Xi_{(x,\tau)}$ in our example so that 
the occupation numbers $n(x,\tau)$ of points on the cluster are changed from 0 to 1 and vice
versa:
\bea
&&\Xi_{(x,\tau)}|1>=|0>\nonumber\\
&&\Xi_{(x,\tau)}|0>=|1>\nonumber
\eea
The clusters in $\macros(v)$ are constructed searching closed loops with the following algorithm.\\
\begin{algo} This algorithm finds a cluster $\cluster$ in $\macros(v)$:
\begin{enumerate}
{\rm
\item To start a loop one first selects a lattice point $(x,\tau)$. 
\item The occupation number $n(x,\tau)$ participates in two plaquettes, 
one before and one after $\tau$. For
$n(x,\tau) = 1$ we consider the plaquette at the later time and for $n(x,\tau) = 0$ we consider the plaquette at the earlier time. 
\item The corresponding plaquette 
is characterized by the occupation numbers of four 
points in $\Lambda\times\qint$. 
One of these points will be the next point on the loop.
\begin{itemize}
\item For a plaquette $[0,0,0,0]$ or $[1,1,1,1]$
the next point is with probability $p_1$
the time-like nearest neighbor of $(x,\tau)$,
and with probability $1 - p_1$ the next-to-nearest
(diagonal) neighbor of $(x,\tau)$ on the plaquette.
\item For a plaquette $[0,1,0,1]$ or $[1,0,1,0]$
the next point on the loop is
with probability $p_2$ the time-like nearest
neighbor, and with probability $1 - p_2$ the
space-like nearest neighbor of $(x,\tau)$. 
\item For a plaquette
$[0,1,1,0]$ or $[1,0,0,1]$ the next point is with probability
$p_3$ the diagonal neighbor, and with
probability $1 - p_3$ the space-like
nearest neighbor of $(x,\tau)$.
\end{itemize} 
\item Once the next point on the loop is
determined the process is repeated from 2 until the loop closes.
\item The points on the closed loop determine the cluster $\cluster$. 
}
\end{enumerate}
\label{alg3}
\end{algo}
The sets of all clusters determined by this algorithm for some arbitrary values of $p_1,p_2,p_3$ is only a subset of the macros $\macros(v)$. This subset is, however, sufficient for the construction of our Monte Carlo algorithm.\\

If the start point in algorithm \ref{alg3} is chosen randomly and the probabilities $p_1,p_2$ and $p_3$ are chosen in the following way
\be
\left(\begin{array}{c}
p_1\\p_2\\p_3
\end{array}\right)= 
\left(\begin{array}{c}
\frac{1}{2}(1+e^{-\beta\epsilon})\\
\frac{1}{2}(1+e^{-\beta\epsilon})/\cosh(\beta\epsilon)\\
(1-\frac{1}{2}(1+e^{-\beta\epsilon}))/\sinh(\beta\epsilon)
\end{array}\right)\label{prob}
\ee
an update of $|v>$ realized by a flip $\Xi_\cluster$ 
obeys detailed balance \cite{wiese}.
The part of the weights proportional to $\exp(-\beta\epsilon(1 - \frac{\mu}{4}))$
is taken into account by a global Metropolis step. 
In the Metropolis step
the cluster is flipped with probability
$p = \mbox{min}(1,\exp(\beta (4 - \mu) W(\cluster)))$
where $W(\cluster) = \frac{1}{4T} \sum_{(x,\tau) 
\in {\cluster}} (2n(x,\tau) - 1)$. This defines the loop algorithm $\phi_{|w|}$.\\

The search method of the clusters in algorithm 
\ref{alg} can be performed using algorithm \ref{alg3} choosing the starting point $(x,\tau)$ from a list of points in $\Lambda\times \qint$ and a list of random numbers.  There, the choice of the probabilities $p_1,p_2$ and $p_3$ is arbitrary. However, to
 obtain a good acceptance $<P_A>$ it is convenient to use the probabilities defined in (\ref{prob}). The weight $\tilde w(v)$ is defined by algorithm \ref{alg} and eq. (\ref{w2}). For our runs we have used a hashing table with 10000 entries. The loop algo
rithm $\phi_{|w|}$ applied in algorithm \ref{alg2} defines our Monte Carlo algorithm $\phi_{|\tilde w|}$. In our simulation we have not found any collision in the hashing table so that the systematic error $\delta$ on the observables is zero.\\

The model in our example is trivial and
can be solved in momentum space. This allows us to test our algorithm. 
By introducing
$c_p^* = \frac{1}{L} \sum_x \exp(ipx) c_x^*$,
$c_p = \frac{1}{L} \sum_x \exp(-ipx) c_x$,
the Hamiltonian in momentum space becomes
$H = \sum_p \hat{p}^2 c_p^* c_p$ 
with $\hat{p}_i = 2 \sin(p_i/2)$. In the grand canonical ensemble
the expectation value of the occupation number is given by
\be
\langle n_x \rangle = \frac{1}{Z} \mbox{Tr} [n_x \exp(- \beta (H - \mu N))] =
\frac{1}{L^2} \sum_p \frac{1}{1 + \exp(\beta (\hat{p}^2 - \mu))}\label{exsol}
\ee

In Table 1 we present the results of the Monte Carlo simulations performed with the loop algorithm and with our algorithm \ref{alg2} and we compare the obtained results with the exact solution (\ref{exsol}). 
We have applied the algorithms for various values of $\beta$ and $\mu$ at fixed lattice spacing $\beta\epsilon=1/16$. The results of both algorithms agree with the exact results within the error bars. It is evident that the sign problem becomes severe for
 the loop algorithm when the temperature is lowered or the chemical potential is increased. However, the sign remains always positive for our algorithm. \\
 
\section{Conclusion}
We have presented a new algorithm for significantly improving quantum Monte Carlo simulations of models plagued by the negative sign problem. Its complexity is only linear in the volume of the lattice used for the simulation.
The generality of this algorithm allows us to apply it to any quantum spin system.  The efficiency of this algorithm was tested on a simple fermionic model. In this example the sign problem is solved.
A more exhaustive analysis of the dynamics of the algorithm is under study. Applications of it to more physically interesting models are also planned.\\

{\Large {\bf Acknowledgments}}\\

I would like to especially thank N. Galli for discussions and help. 
I also would like to thank B. Jegerlehner for helpful comments and P.Weisz for reading the manuscript and helpful comments.


\begin{table}[p]
\begin{center}
{\footnotesize
\begin{tabular}{|c|c|c|c|c|c|c|c|c|c|}\hline
$\beta$&$\mu$&$\sqrt{|\Lambda|}$&$4T$&$<n>_O$&$<n>_L$&$<n>_E$&$<sgn>_O$&$<sgn>_L$& $<P_A>$\\\hline\hline
0.5 & 2 & 8 & 32 & 0.304(2) & 0.305(3) & 0.3049 & 1.0(0) & 0.690(4) & 0.64(1) \\\hline
0.5 & 4 & 8 & 32 & 0.501(2) & 0.498(3) & 0.5000 & 1.0(0) & 0.591(4) & 0.64(1) \\\hline
1.0 & 2 & 8 & 64 & 0.233(2) & 0.230(40) & 0.2321 & 1.0(0) & 0.048(6) & 0.612(9)\\\hline
1.0 & 3 & 8 & 64 & 0.357(2) & ? & 0.3568 & 1.0(0) & -0.003(6) & 0.60(1)\\\hline
1.0 & 4 & 8 & 64 & 0.499(2) & ? & 0.5000 & 1.0(0) & 0.002(8) & 0.59(1)\\\hline
2.0 & 2 & 8 & 128 & 0.193(2) & ? & 0.1956 & 1.0(0) & 0.004(9) & 0.58(1)\\\hline
2.0 & 4 & 8 & 128 & 0.498(2) & ? & 0.5000 & 1.0(0) & -0.001(9) & 0.58(1)\\\hline\hline
1.0 & 0 & 12 & 64 & 0.067(2) & 0.069(2) & 0.0685 & 1.0(0) & 0.703(4) & 0.64(1)\\\hline
1.0 & 1 & 12 & 64 & 0.134(2) & 0.139(5) & 0.1351 & 1.0(0) & 0.295(8) & 0.61(1)\\\hline
1.0 & 2 & 12 & 64 & 0.232(2) & 0.20(10) &  0.2321 & 1.0(0) & 0.018(6) & 0.60(1)\\\hline
1.0 & 3 & 12 & 64 & 0.358(2) & ? & 0.3568 & 1.0(0) & 0.001(3) & 0.59(1)\\\hline
1.0 & 4 & 12 & 64 & 0.501(2) & ? & 0.5000 & 1.0(0) & -0.001(4) & 0.58(1)\\\hline
1.0 & 5 & 12 & 64 & 0.642(2) & ? & 0.6432 & 1.0(0) & -0.002(5) & 0.59(1)\\\hline
1.0 & 6 & 12 & 64 & 0.768(2) & ? & 0.7679 & 1.0(0) & 0.001(4) & 0.58(1)\\\hline
1.0 & 7 & 12 & 64 & 0.863(2) & ? & 0.8649 & 1.0(0) & -0.001(4) & 0.55(1)\\\hline\hline
\end{tabular}}
\end{center}
\caption{{\small Results from the Monte Carlo simulations 
for different parameters. The simulations are done with
the loop algorithm $\phi_{|w|}$ and our algorithm \ref{alg2}. 
The expectation values of the number operator $n$ and the sign obtained with the loop algorithm are denoted by $<n>_L$ and $<sgn>_L$, the ones 
obtained with our algorithm \ref{alg2} are denoted by $<n>_{O}$ and $<sgn>_O$. The mark "?" indicates that the measurement of the expectation value was impossible, because the statistical error exceeds the expectation value. The exact values of $<n>$ are 
denoted by $<n>_E$. 
The average of the acceptance $<P_A>$ of the algorithm \ref{alg2} is also measured.}}
\end{table}

\end{document}